\documentclass[epj]{svjour}
\usepackage{graphics}
\usepackage[isolatin]{inputenc}
\usepackage{dcolumn}
\begin{document}
\title{Precise atomic masses of neutron-rich Br and Rb nuclei close to the r-process path}
\author{S. Rahaman\thanks{\emph{E-mail address:} saidur.rahaman@phys.jyu.fi}, U. Hager, V.-V. Elomaa, T. Eronen, J. Hakala, A. Jokinen, A. Kankainen, 
P. Karvonen, I.D.~Moore, H. Penttilä, S. Rinta-Antila, J. Rissanen, A. Saastamoinen, T. Sonoda and J. Äystö 
%
}                     
\offprints{}          
\institute{Department of Physics, P.O. Box 35 (YFL), FIN-40014 University of Jyväskylä, Finland}
\date{Received: date / Revised version: date}
%
\abstract{ The Penning trap mass spectrometer JYFLTRAP, coupled to the Ion-Guide Isotope Separator On-Line (IGISOL) facility at Jyväskylä, was 
employed to measure the atomic masses of neutron-rich $^{85-92}$Br and $^{94-97}$Rb isotopes with a typical accuracy less than 10 keV. Discrepancies 
with the older data are discussed. Comparison to different mass models is presented. Details of nuclear structure, shell and subshell closures are 
investigated by studying the two-neutron separation energy and the shell gap energy. 
} 
\authorrunning{S. Rahaman, et al.}
\maketitle
\section{Introduction}
\label{intro}
Nuclei with masses \emph{A} $\sim$ 100 have received a considerable interest in recent years both theoretically as well as experimentally. Reasons for 
this interest stem from shell closure effects at \emph{N} = 50, the subshell closure at \emph{N} = 56 and an onset of large deformation for the nuclei 
with \emph{N} $\geq$ 60. A systematic study of the properties of a series of neutron-rich isotopes of rubidium, strontium, zirconium and molybdenum 
reveals a rapidly changing behavior in nuclear structure in the region of neutron number 58 $\leq$ \emph{N} $\leq$ 61. This behavior has been studied 
experimentally using several methods employing $\beta$-decay \cite{SRA07}, $\gamma$-ray measurements \cite{HBD06}, collinear laser spectroscopy 
\cite{PC02} and most recently direct mass measurements \cite{UH06}. The deformation effect is dominant in the strontium and zirconium isotope chains 
where the onset of deformation is characterized by a breaking of a smooth trend of two-neutron separation energies. 

Systematic theoretical calculations employing the Hart\-ree Fock-Bogoliubov (HFB) model predict that the strontium, zirconium and molybdenum nuclei 
with \emph{N} $\leq$ 58 are nearly spherical in shape whereas those with \emph{N} $\sim$ 60 prefer a highly deformed prolate shape \cite{AK85}. The 
JYFLTRAP precise experimental observations indicate this occurrence \cite{UH06}. Extension of this knowledge of deformation changes towards the lower 
\emph{Z} region is of particular interest in the present study of rubidium and bromine isotopes.

\begin{figure}
\resizebox{0.5\textwidth}{!}{%
  \includegraphics{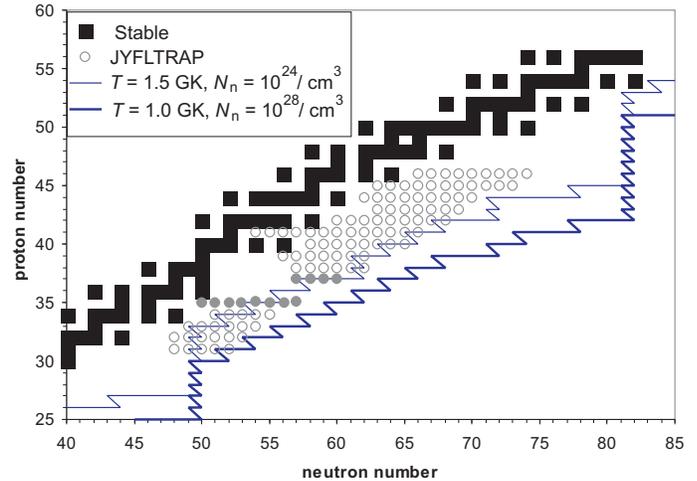}
}
\caption{The  neutron-rich nuclei (empty circles) investigated at JYFLTRAP and the two r-process paths \cite{PA65,VB96} at two different temperatures 
and neutron densities. The rubidium and bromine isotopes studied in this article are indicated by filled grey circles.}
\label{fig:1}       
\end{figure}

Additional motivation for this work derives from the rapid neutron capture process (r-process) on heavy elements. The r-process is considered to be 
responsible for the synthesis of the heavy elements beyond iron (\emph{Z} $\geq$ 26) and approximately half of all nuclear species in nature are 
produced via this process. In a good approximation the neutron capture reactions proceed in a (n, $\gamma$) $\leftrightarrow $ ($\gamma$, n) 
equilibrium, fixing the r-process reaction path at rather low neutron separation energies of about 2 to 3 MeV \cite{KL03}.             

The majority of nuclei in the r-process path are not accessible experimentally, therefore the theoretical calculations involve extensive 
extrapolations in the direction of heavier and more neutron-rich isotopes. For reliable nucleosynthesis calculations, nuclear masses (or the neutron 
separation energy $S_{n}$), half-lives and beta-decay \emph{Q} values are important parameters \cite{JC91}. To first order, the values of $S_{n}$ 
determine the r-process path and the values of the half-lives determine the shape of the abundance curve. Thus the nuclear mass is one of the 
important nuclear parameters for the r-process calculations.

The study of nuclear properties via precision atomic mass measurements of exotic nuclei is an important part of the JYFLTRAP setup at the Ion-Guide 
Isotope Separator On-Line (IGISOL) facility \cite{JA01} in Jyväskylä. So far, neutron-rich isotopes of Ga, Ge, As, Se, Br, Rb, Sr, Y, Zr, Mo, Tc, Ru, 
Rh and Pd have been investigated \cite{UH06,AJ06,AJ07,JH07,UH06a}. Two r-process paths calculated by using the canonical model \cite{PA65,VB96} and 
the neutron-rich nuclei investigated at JYFLTRAP are shown in Fig.~1. Some of the investigated nuclei at JYFLTRAP are directly recline on the 
r-process paths. In this paper experiments on neutron-rich Rb and Br isotopes around the mass region \emph{A} $\sim$ 100 will be presented and 
discussed. The masses of Br and Rb isotopic chains were previously measured employing $\beta$-decay endpoint energies and using a dipole mass 
spectrometer, respectively. Recently ISOLTRAP \cite{PA06} has also measured masses of neutron-rich Kr isotopes up to $^{95}$Kr. 
\section{Experimental setup and procedure}
\label{sec:1}

\subsection{JYFLTRAP}
\label{sec:2}


JYFLTRAP \cite{VK04} is an ion trap experiment for cooling, bunching, isobaric purification and precision mass measurements of radioactive ions 
produced at the IGISOL facility. In this work radioactive nuclides were produced in a proton-induced fission reaction by bombarding a thin (15 
mg/cm$^{2}$) natural uranium target with a 30 MeV proton beam of average intensity of 7 $\mu$A from the Jyväskylä K-130 cyclotron. The production 
rates of the studied nuclides varied from 10,000 ions/s to 100 ions/s for the most exotic isotope. The ions were extracted from the gas cell by helium 
gas flow into a differential pumping stage where they were accelerated to 30 keV and mass-separated with a 55$^{0}$ dipole magnet.  The ions were 
transported to the JYFLTRAP setup where they were efficiently cooled and cleaned from possible isobaric contaminants before their mass was measured. 
To perform this process JYFLTRAP consists of two main components as shown in fig.~2, a helium-filled RFQ cooler \& buncher \cite{AJ02,AN01,JA03} and 
two cylindrical Penning traps in one 7 T superconducting magnet.

%
\begin{figure}
\resizebox{0.50\textwidth}{!}{%
\includegraphics{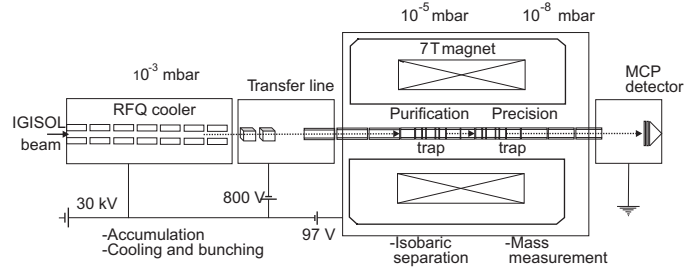}
}
\caption{A schematic drawing of the JYFLTRAP setup. The main components are the RFQ cooler \& buncher and two cylindrical Penning traps. The Penning 
traps are inside a single 7 T superconducting solenoid. A microchannel plate (MCP) detector is placed at the end of the setup which is at ground 
potential.}
\label{fig:2}       
\end{figure}

The ions delivered to the RFQ cooler \& buncher were cooled, accumulated and finally bunched at a low kinetic energy with an  emittance of 3 
$\pi$-mm-mrad (measured at 38 keV) \cite{AN03}. These ion bunches were transported to the Penning trap system. The traps are placed in the homogeneous 
regions of the magnet separated by 20 cm. The first trap, called the purification trap is filled with helium gas at a pressure of about 10$^{-5}$ mbar 
and is dedicated to isobaric cleaning using a mass selective buffer gas cooling technique \cite{GS91}. The second trap, the precision trap has a 
residual pressure of about 10$^{-8}$ mbar. It is used for high precision mass measurements using the time-of-flight ion cyclotron resonance technique 
\cite{MK95}. 
   
The cooled and bunched ions from the RFQ buncher were captured in the purification Penning trap and further cooled by the collisions with helium 
atoms, finally being stored at the axial potential mimima of the trap. Possible isobaric contaminations were removed by applying successive RF dipole 
and quadrupole excitations in the presence of the helium gas. As a result, the ions of interest were centered when the applied quadrupole excitation 
frequency was close to the ion cyclotron frequency. A cycle time of about 400 ms was used in the purification trap and a mass resolving power of about 
10$^{5}$ was reached with this configuration.

  The isobarically purified and cooled ions were transported to the precision trap. RF dipole excitation at the ion's magnetron frequency ($\nu_{-}$) 
was applied in order to enhance the magnetron radius employing the phase-locking technique \cite{KB03}. The RF quadrupole excitation was then applied 
at the sum frequency $\nu_{c}=\nu_{+} + \nu_{-}$ to convert the slow magnetron motion to the reduced cyclotron motion with a higher radial energy. 
$\nu_{c}$ and $\nu_{+}$ are the cyclotron and reduced cyclotron frequency of the ions respectively. The excitation time varied from 150 ms to 800 ms 
(see table~1). Finally, the ions were ejected from the trap. They drift towards the microchannel plate (MCP) detector through a set of drift tubes 
where a strong magnetic field gradient exists. Here the ions experience a force which converts the radial kinetic energy of the ions into longitudinal 
energy additional to the 30 keV acceleration from the high-voltage platform to the ground potential where the MCP is situated. As a result an ion 
which was in resonance with the applied excitation quadrupole field moves faster towards the detector than an ion which was off resonance. By 
measuring the time of flight as a function of the applied excitation frequency, the cyclotron frequency $\nu_{c}$ was determined. A typical 
time-of-flight resonance spectrum is shown in fig.~3. This spectrum yields the ion cyclotron frequency

\begin{figure}
\resizebox{0.5\textwidth}{!}{%
  \includegraphics{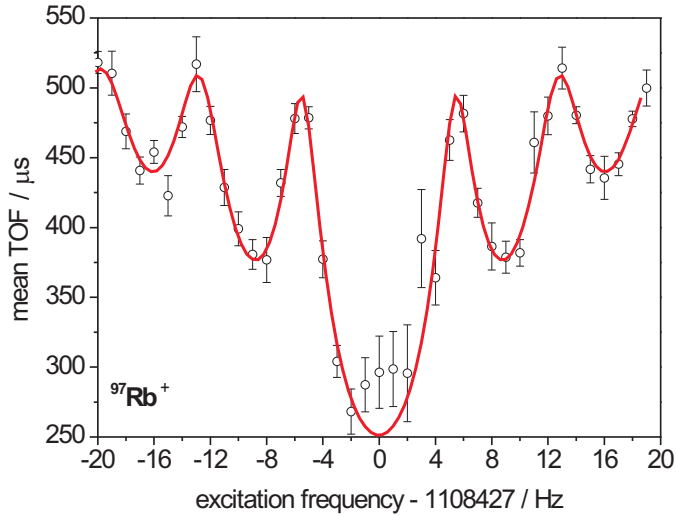}
}
\caption{Time-of-flight (TOF) resonance of $^{97}$Rb$^{+}$ ions measured at JYFLTRAP. The line shows a fit to the data points using the theoretically 
expected line shape. The excitation time used for this isotope is 150 ms.}
\label{fig:3}       
\end{figure}

\begin{equation}
\nu_{c} = \frac{1}{2\pi}\frac{q}{m}B,
\end{equation}
where \emph{B} is the magnetic field, \emph{m} is the mass and \emph{q} the charge state of the ion. To calibrate the magnetic field, the cyclotron 
frequency ($\nu_{c}^{ref}$) of a precisely known reference mass ($m_{ref}$) was measured. The atomic mass of the ion of interest was then determined 
using the equation 
\begin{equation}
m = r \cdot(m_{ref} - m_{e}) + m_{e},
\end{equation}
where r = $\nu_{c}^{ref}$/$\nu_{c}$ is the frequency ratio and $m_{e}$ is the electron mass. 
 
%
\begin{table}
\caption{The excitation time used for Br and Rb isotopes for the mass measurements is indicated. $^{88}$Rb was used as a reference ion. Measurements 
were performed in July 2005.}
\label{tab:1}       
\begin{tabular}{lll}
\hline\noalign{\smallskip}
Measured isotopes  & Excitation time (ms)  \\
\noalign{\smallskip}\hline\noalign{\smallskip}
$^{85-90}$Br & 800  \\
$^{91}$Br  & 400  \\
$^{92}$Br  & 150  \\
$^{94}$Rb & 800 \\
$^{95}$Rb  & 300 \\
$^{96-97}$Rb & 150  \\
\noalign{\smallskip}\hline
\end{tabular}
\end{table}

\subsection{Evaluation of uncertainties}
\label{sec:3}

The dominating systematic uncertainties in the cyclotron frequency determination are the uncertainties due to magnetic field fluctuations and the mass 
difference between the reference ion and the ion of interest. A shift of the center value of the cyclotron frequency due to contaminating ions in the 
trap is a further possible source of systematic error. 
             
The drift of the magnetic field was taken into account by the linear interpolation of the reference cyclotron frequencies measured before and after 
the cyclotron frequency measurement of the ion of interest. As the measurement time increases the reliability of this linear interpolation decreases 
because of the short term magnetic field fluctuation due to external influences as for example, temperature, pressure and the presence of 
ferromagnetic materials near the superconducting magnet \cite{AK03}. 30 minutes of accumulation time for each file ensure the reliability of the 
measurement. The cyclotron frequency of the reference ions was plotted as a function of time. Two consecutive reference frequencies were not deviated 
by each other more than 50 mHz. Experimental study showed that this fluctuation is random and a deviation of 50 mHz was derived from the plot. This 
uncertainty was added quadratically to the statistical uncertainty of each measured frequency. A mass dependent systematic error on the order of 
$10^{-10}$/u was derived from the off-line measurement by comparing the frequencies of $^{129}$Xe and O$_{2}$ ions. This uncertainty was added to the 
final average frequency ratio of each nuclei.

If the ions stored in the trap have equal mass, the driving field will act on the center of mass of all ions and there will be no frequency shift 
\cite{GB92}. However, some of the investigated nuclei had very short half-lives, for example $^{97}$Rb ($T_{1/2}$ = 170 ms). Therefore the decay 
products and the ions created by charge exchange processes were a source of the contaminating ions. In order to correct for this the center value of 
the cyclotron frequency is plotted as a function of the number of detected ions and the data is fitted by a linear function and extrapolated to the 
observed 0.6 ion (detector efficiency 60 \%), equivalent to one ion in the trap \cite{AK03}. The detected number of ions was kept below 2 to 3 ions 
per bunch to minimize this shift. One example, in the case of the $^{96}$Rb measurement the amount of cyclotron frequency shift was 3.6 mHz/ion. In 
the case of the other nuclei the size of the frequency shift were varied from 0-8 mHz /ion. The uncertainties at the extrapolated point (0.6 ion) were 
found to be a factor of two times larger than at 2 ions per bunch. In the cases where the ion class analysis was not possible due to the lower number 
of ions, the uncertainty of the frequency was increased manually to an average value. This average value was determined from the cases in which the 
ion class analysis was applied successfully.
   
\section{Results}
\label{sec:2}

The results of the mass measurements for $^{85-92}$Br and $^{94-97}$Rb performed at JYFLTRAP are summarized in table~2 and discussed in sections 3.1 
and 3.2. The cyclotron frequency ratio of each isotope is given with respect to the cyclotron frequency of the $^{88}$Rb reference isotope. The mass 
of $^{88}$Rb is known with an uncertainty of 160 eV \cite{GA03}. The uncertainty of all the investigated isotopes reached a level below 10 keV. In 
table~2 the column ME$_{ex}$ presents the measured JYFLTRAP mass excess with the final uncertainty. The uncertainty contains the statistical and all 
known systematic uncertainties. 
All previous measurements of Br masses were extracted from the endpoint energy of the beta spectrum. In the case of the Rb isotopes a dipole mass 
spectrometer and a Penning trap were used up to $^{94}$Rb. For heavier Rb isotopes masses were previously measured up to $^{99}$Rb but with moderate 
precision only.

\begin{table*}
\begin{center}
\caption{Results from the analysis of $^{85-92}$Br and
$^{94-97}$Rb measured at JYFLTRAP. The measured average frequency ratio
$\overline{r}$ and its uncertainty is presented. $T_{1/2}$ represents the beta decay half-life of the studied nuclei. ME$_{ex}$ represents the 
experimental mass excess obtained from the cyclotron frequency ratio. ME$_{lit}$ are the AME03 values \cite{GA03}. The last column gives the 
difference between the AME03 and JYFLTRAP values $\Delta$ = ME$_{lit}$ - ME$_{ex}$.}
\label{tab:2}       
\begin{tabular}{llllll}
\hline\noalign{\smallskip}
Nucleus&$T_{1/2}$&Freq. ratio$(\overline{r})$ = $\frac{\nu_{c}^{ref}}{\nu_{c}}$&ME$_{ex}$ / keV &ME$_{lit}$ / keV & $\Delta$ / keV \\
\noalign{\smallskip}\hline\noalign{\smallskip}
$^{85}$Br & 2.87 min&0.965 237 440(37)   & -78575.4(35)& -78610(19)&-34.6  \\
$^{86}$Br &55.10 sec&0.977 334 856(38)  & -75632.3(35)&-75640(11) & -7.7\\
  $^{87}$Br &55.70 sec&0.988 731 283(39)  &-73892(4) &-73857(18) &35 \\
 $^{88}$Br &16.30 sec&1.000 145 235(39)  &-70716(4) & -70730(40)&-14 \\
  $^{89}$Br &4.40 sec&1.011 550 223(40)  &-68275(4) & -68570(60)&-295 \\
  $^{90}$Br &1.90 sec&1.022 977 586(41)  &-64001(4) &-64620(80) &-619 \\
  $^{91}$Br & 640 ms&1.034 388 086(43)  & -61108(4)&-61510(70) &-402 \\
 $^{92}$Br & 343 ms&1.045 822 783(82)  &-56233(7) &-56580(50) &-347 \\
\hline
 $^{94}$Rb &2.69 sec&1.068 422 547(45) &-68564(5) &-68553(8) &11 \\
  $^{95}$Rb &377 ms&1.079 829 822(38) & -65935(4)&-65854(21) &81\\
 $^{96}$Rb &199 ms&1.091 260 919(41)  & -61355(4)&-61225(29) &130\\
 $^{97}$Rb & 170 ms&1.102 670 725(65)  &-58519(6) &-58360(30) &159\\
\noalign{\smallskip}\hline
\end{tabular}
\end{center}
\end{table*}

\subsection{$^{85-92}$Br isotopes}
\label{sec:1}

\subsubsection{$^{85}$Br}
Earlier, the mass of $^{85}$Br has been primarily derived from the beta-decay study of K. Aleklett \emph{et al.} \cite{KA79}. A beta spectrum yielded 
Q$_\beta$ = 2870(19) keV and a mass excess of -78600(20) keV. The adopted value in the AME03 is -78610(19) keV. The JYFLTRAP value -78575.4(35) keV 
differs by 35 keV or 2 standard deviations from the adopted value. 
\subsubsection{$^{86}$Br}
The mass of this nucleus was deduced from its beta decay studied by K. Aleklett \emph{et al.} \cite{KA79}. A Q$_\beta$ value of 7620(60) keV and a 
mass excess value -75640(60) keV were derived from the beta spectrum gated by the 2751 keV transition which depopulates the level at 4316 keV in  
$^{86}$Kr. M. Groß \emph{et al.} \cite{MGross92} also studied this nucleus in 1992 and extracted a  mass excess of -75636(12) keV. The JYFLTRAP value 
-75632.3(35) keV is in good agreement with the value of -75640(11) keV adopted in AME03.    
\subsubsection{$^{87}$Br}
The previous mass excess value of $^{87}$Br has been extracted from the beta endpoint energy \cite{KA79,MGross92}. The JYFLTRAP value -73892(4) keV  
differs by 2 standard deviations from the value of -73857(18) keV adopted in the AME03 mass table.

\begin{table}
\caption{The \emph{Q}-value of the Br isotopes from the Penning mass measurements (this work and \cite{PA06}) is indicated.}
\label{tab:1}       
\begin{tabular}{lllll}
\hline\noalign{\smallskip}
isotopes & \emph{Q}-value from trap& \emph{Q}-value & difference \\
 &  mass measurement  & from AME03  & \emph{Q}$_{trap}$-\emph{Q}$_{AME03}$\\
  & \emph{Q}$_{trap}$ (keV)  & \emph{Q}$_{AME03}$ (keV)& (keV) \\
\noalign{\smallskip}\hline\noalign{\smallskip}
$^{85}$Br & 2905(4) &2870(19)&-35 \\
$^{86}$Br & 7632(4) &7626(11)&-6\\
$^{87}$Br & 6817(5)   &6852(18)&35\\
$^{88}$Br & 8975(5) &8962(40)&-13\\
$^{89}$Br  & 8261(4) &8160(30)&-101\\
$^{90}$Br & 10959(4) &10350(80)&-609 \\
$^{91}$Br  &9867(5) &9800(40)&-67\\
$^{92}$Br & 12536(8) &12205(50)& -331\\
\noalign{\smallskip}\hline
\end{tabular}
\end{table}

\begin{figure}
\resizebox{0.5\textwidth}{!}{%
  \includegraphics{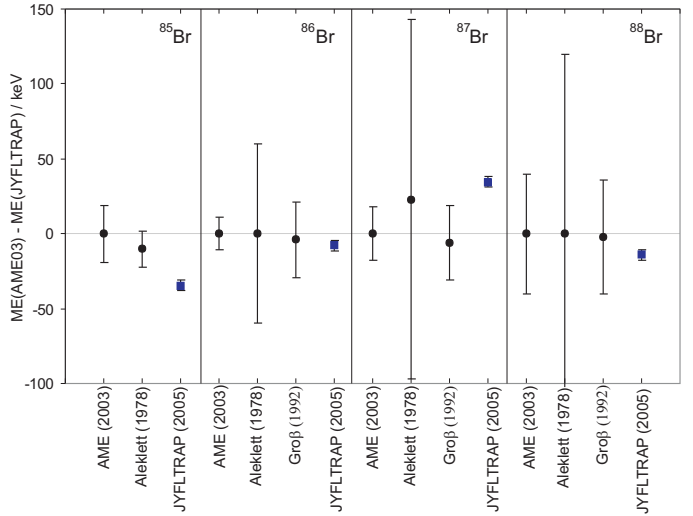}
}
\caption{Experimental mass excess of $^{85-88}$Br isotopes relative to the AME03 values. The filled circles indicate beta-decay experiments of K. 
Aleklett \emph{et al}. (1979) \cite{KA79} and M. Groß \emph{et al}. (1992) \cite{MGross92}. The filled squares are the values from JYFLTRAP.}
\label{fig:4}       
\end{figure}

\begin{figure}
\resizebox{0.5\textwidth}{!}{%
  \includegraphics{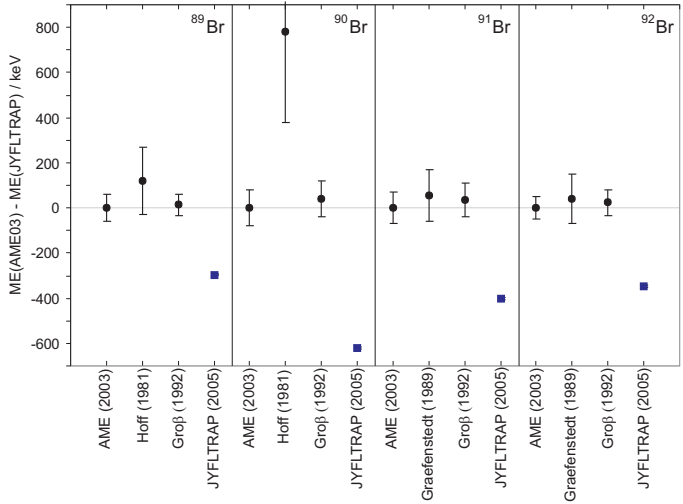}
}
\caption{Experimental mass
excess of $^{89-92}$Br isotopes relative to the AME03 values. The filled circles indicate beta-decay experiments of P. Hoff \emph{et al}. (1981) 
\cite{PH81}, M. Graefenstedt \emph{et al}. (1989) \cite{MG89} and M. Groß \emph{et al.} (1992) \cite{MGross92}. The filled squares are the values from 
JYFLTRAP.}
\label{fig:5}       
\end{figure}

\subsubsection{$^{88}$Br}
The endpoint energy of a beta spectrum has been determined in coincidence with cascade $\gamma$ transitions to the ground state of $^{88}$Kr 
\cite{KA79}. An average Q$_\beta$ = 8970(130) keV and a mass excess value of -70730(120) keV was derived. It was also investigated by M. Groß \emph{et 
al.} \cite{MGross92} yielding almost the same mass excess value. The JYFLTRAP value -70716(4) keV is in good agreement with the previous values. 
$^{88}$Br has an isomeric state at an energy of 272.7 keV with a half life of 5.4 $\mu$s. The ground state mass measurement will not be affected by 
such a short-lived isomeric state since it will decay before reaching the Penning trap system. 
 
\subsubsection{$^{89}$Br}
Spectroscopic investigations of $^{89}$Br nuclei have been performed extensively by P. Hoff \emph{et al.} \cite{PH81} in 1981 using a molecular ion 
$^{27}$Al$^{89}$Br$^{+}$. The endpoint energy was derived from the observed beta spectra in coincidence with selected $\gamma$ transitions in 
$^{89}$Kr having an energy more than 4 MeV and possessing strong $\beta$ branches. An average Q$_\beta$ = 8140(140) keV leads to a mass excess value 
of -68690(150) keV. M. Groß \emph{et al.} \cite{MGross92} also investigated the beta endpoint energy of $^{89}$Br. Our measured value -68275(4) keV 
differs by 295 keV or 5 standard deviations from the value adopted in AME03. This deviation is reduced to 100 keV taking into account the $^{89}$Kr 
mass value measured recently at ISOLTRAP\cite{PA06} (see table 3).
            
\subsubsection{$^{90}$Br}
The mass excess value of $^{90}$Br was determined from the endpoint energy of a beta spectrum by P. Hoff \emph{et al.} \cite{PH81}. The beta spectrum 
observed in coincidence with the 3231 keV $\gamma$ transition of $^{90}$Kr yielded an average Q$_\beta$ of 9800(400) keV and the mass excess of 
-65400(400) keV. Another experiment by M. Groß \emph{et al.} on the beta endpoint energy provided a mass excess value of -64660(80) keV. The JYFLTRAP 
value -64001(4) keV differs by 619 keV from the value adopted in the AME03 Mass Table. The difference of nearly 8 standard deviations can be due  to 
various reasons. Firstly, a low statistics beta spectrum can easily yield an MeV error in an endpoint energy determination. Secondly, the energy 
levels of $^{90}$Kr are poor known. Recently, the measured mass value of the $^{90}$Kr did not improve the observed deviation as indicated in table 3.       

\subsubsection{$^{91}$Br}
The mass value of $^{91}$Br has been determined at the ISOLDE II mass separator by studying the end-point energy of the beta spectrum measured by M. 
Graefenstedt \emph{et al.} in 1989 \cite{MG89}. The beta spectrum observed in coincidence with a set of gated $\gamma$ transitions was obtained giving 
an average Q$_\beta$ of 9790(100) keV and a mass value of -61565(115) keV. M. Groß \emph{et al.} deduced a mass excess value of -61545(75) keV from 
the beta endpoint energy. The JYFLTRAP value, -61108(4) keV differs from this value by as much as 400 keV. Considering the new mass value of the 
$^{91}$Kr this deviation is reduced to 67 keV tabulated in table 3.    

\subsubsection{$^{92}$Br}
Our measured mass excess value of -56233(7) keV differs from the AME03 value by 350 keV. The AME03 value is based on the previous measurements by M. 
Graefenstedt \emph{et al.} in 1989 \cite{MG89} and M. Groß \emph{et al.} in 1992 \cite{MGross92}. The large difference of 7 standard deviations cannot 
be understood but could have similar origins as discussed before for $^{90}$Br.

 \begin{figure}
\resizebox{0.5\textwidth}{!}{%
  \includegraphics{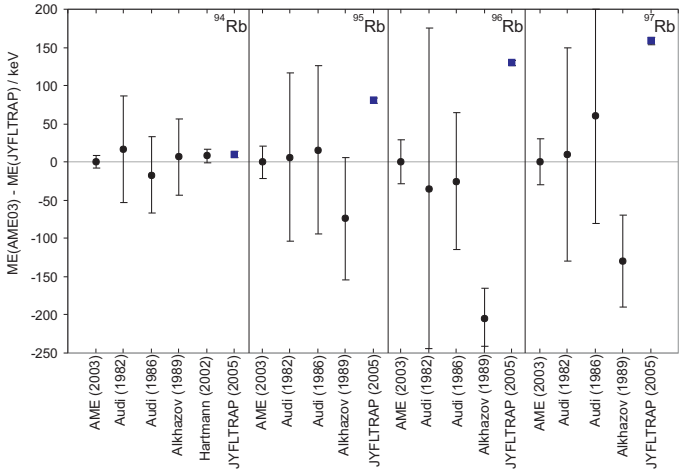}
}
\caption{Experimental mass
excess of Rb isotopes relative to the AME03 values. The filled circles indicate the double focusing spectrometer experiment of G. Audi \emph{et al.} 
(1982) \cite{GA82} and (1986) \cite{GA86}, the ISOLTRAP experiment of H. Raimbault-Hartmann \emph{et al.} (2002) \cite{RH02} and the dipole mass 
spectrometer. of G.D. Alkhazov et at. (1989) \cite{GDA89}. The filled squares are the values from JYFLTRAP.}
\label{fig:6}       
\end{figure}
 
\subsection{$^{94-97}$Rb isotopes}
\label{sec:1}

\subsubsection{$^{94}$Rb}

The mass of $^{94}$Rb was measured with the Orsay double-focusing mass spectrometer in 1982 and again in 1986 with better accuracy \cite{GA82,GA86}. 
In the 1982 run inconsistencies arose in the evaluation of $^{94}$Rb. This finally lead to the suggestion of the existence of two short-lived isomers, 
however this was not confirmed by any other experiment. Finally, it was confirmed that the $^{94}$Rb beam was contaminated. Other direct mass 
measurements took place at the Prism mass spectrometer (PMC) consistent with the Orsay measurements \cite{GDA89}. The mass of $^{94}$Rb was also 
determined at ISOLDE using the Penning trap mass spectrometer with an uncertainty of 9 keV \cite{RH02}. The JYFLTRAP value -68564(5) keV is in 
excellent agreement with all previous measurements.   
      
\subsubsection{$^{95}$Rb}
In our experiment the mass excess of $^{95}$Rb was measured to be -65935(4) keV. The value agrees within the uncertainties with the 1982 and 1986 
measurements of the Orsay groups. The new value differs slightly from the adopted value in the AME03 and disagrees with G.D. Alkhazov \emph{et al.} by 
150 keV \cite{GDA89}.
 
\subsubsection{$^{96}$Rb}
Our measured mass excess of $^{96}$Rb, -61355(4) keV agrees within the error with the value of G. Audi \emph{et al.} from 1982. However it differs 
significantly from the measurement of G.D. Alkhazov \emph{et al.} and the adopted value in the AME03. $^{96}$Rb seems less bound than the AME03 value 
indicates. An isomeric state in $^{96}$Rb is predicted by AME2003 \cite{GA03} to lie at (0$\pm$200 keV) with a half-life $>$ 1($\pm$200) ms. Since no 
other resonance was observed within a scan range of 200 keV,our measured mass excess value could be either ground state or isomeric state. 
\subsubsection{$^{97}$Rb}
The measured mass excess of $^{97}$Rb, -58519(6) keV lies within the uncertainty of the 1986 value. However it differs from the AME03 value by 159 keV 
corresponding to 5 standard deviations. The AME03 value is influenced also by the measurement of G.D. Alkhazov \emph{et al.}, which differs by 290 keV 
from the JYFLTRAP value.
 
\section{Discussion}
\label{sec:4}

The masses of neutron-rich Br and Rb isotopes have been measured by direct means using the Penning trap technique with accuracies better than 10 keV. 
Discrepancy with the older values as discussed in section 3 and summarized in figs.~4,~5 and 6 indicated deviations from the previous values of up to 
600 keV. The shortest lived isotope measured has a half-life of 170 ms ($^{97}$Rb). The new JYFLTRAP data provides significant improvements in the 
knowledge of experimental masses. In the following we will use these data to highlight possible shell structure effects far from stability by viewing 
the systematics of the two neutron-separation energies and by comparing the results with a few selected mass models. 

\begin{figure}
\resizebox{0.5\textwidth}{!}{%
  \includegraphics{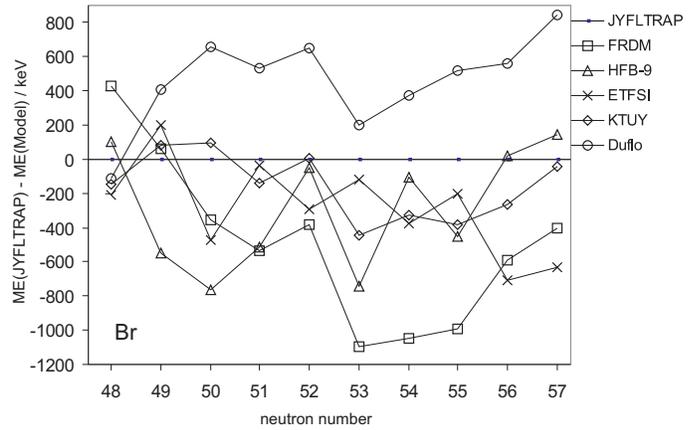}
}
\caption{Comparison of Br mass excess values from JYFLTRAP and calculated values from five leading mass models.}
\label{fig:7}       
\end{figure}

\begin{figure}
\resizebox{0.5\textwidth}{!}{%
\includegraphics{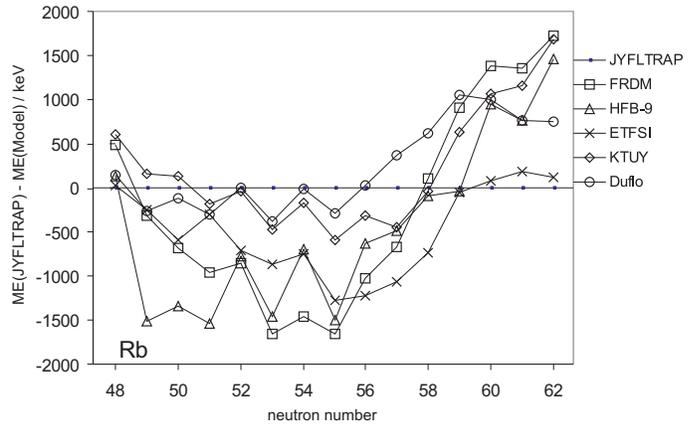}
}
\caption{Comparison of Rb mass excess values from JYFLTRAP and calculated values from five leading models. The AME03 tabulated mass values are used 
for \emph{N} = 48-55 to see an enlarged region.}
\label{fig:8}       
\end{figure}

\subsection{Comparison with mass models}
 \label{sec:1}
It is necessary to compare the models with experimental data for their further development and to improve their predictive power for the region far 
from stability, where mass values are required for example to calculate the r-process path. For comparison, we have chosen five leading nuclear mass 
models generally used in the literature \cite{DL03,JMP04} that have reliably reproduced the known masses.

The Finite-Range Droplet Model (FRDM) by Peter Möller \emph{et al.} \cite{PM95}, belongs to a microscopic and macroscopic (mic-mac) type of 
calculation. This model contains two parts, a macroscopic part (based on the liquid-drop model) and a microscopic part (single-particle model). In the 
macroscopic part the binding energy is a smooth function of proton number \emph{Z} and neutron number \emph{N}, derived from the liquid-drop model. 
Microscopic effects are obtained from a deformed single-particle model. The mass excess values calculated according to this model have the largest 
deviations for both isotope chains investigated in this work as shown in figs.~7 and 8. No significant odd-even staggering is observed in this region. 

The Hartree-Fock Bogoliubov model (HFB-9) \cite{DD84} is classified as fully microscopic, since the model is based on an effective two-body 
(nucleon-nucleon) interaction. This also includes pairing energy. The advantage of this model is that it permits one to derive all properties of the 
nucleus from internucleonic forces. A drawback of this model is that it generally does not succeed in calculating odd nuclei which is an inconvenience 
for astrophysics. The odd-even staggering is clearly overestimated in this model and can be seen for Br and Rb isotopes shown in figs.~7 and 8. 

The Extended Thomas-Fermi and Strutinsky Integral (ETFSI) model \cite{JMP96,YA95} adopted a semi-classical approximation to the Hartree-Fock method 
which includes full Strutinsky shell corrections. They extend Thomas-Fermi calculations to be available for a Skyrme-type (zero-range) interaction and 
a delta function pairing force incorporating a full Strutinsky integral. Mass excess values calculated according to this model are better relative to 
the previous two models for the isotopes studied in this work. Odd-even staggering is observed but less than that of the HFB-9 model.

Another mass  model \cite{HK00} called KTUY also has two parts and belongs to the mic-mac family. The macroscopic part is a smooth function of 
\emph{Z} and \emph{N} based on the liquid-drop model including an average pairing energy. The microscopic part calculates the intrinsic shell energy 
and the average deformation energy of a deformed nucleus employing a single-particle potential. The mass excesses from this model agree reasonably 
well with the JYFLTRAP experimental values both for Br and Rb. For the heaviest Rb isotopes, however, a large difference of nearly 1 MeV is observed. 
Only small odd-even staggering is noticed for \emph{N} = 50-53 in the Br isotope chain and for \emph{N} = 50-57 in the Rb isotope chain. It is of 
interest to note that the KTUY model fails completely in predicting masses of neutron-rich Tc-Pd isotopes as shown by another study performed recently 
at JYFLTRAP \cite{UH06a}.

The Duflo and Zuker mass model \cite{JD95} is more fundamental than the mic-mac approach but it is not yet fully microscopic since no nucleonic 
interaction appears explicitly. This model starts with an assumption that there exists an effective interaction (pseudopotential) which makes the HF 
calculation possible. Finally, it can be written as an effective Hamiltonian which contains two parts, a monopole term and a multipole term. The 
monopole calculations are purely HF-type based on single particle properties while the multipole term acts as a residual interaction and the 
calculation goes beyond HF.
The mass excess for Rb isotopes does not differ significantly for lower neutron number but it starts to differ for higher neutron number. For example 
this model has a 1 MeV discrepancy compared to experiment at \emph{N} = 59. The odd-even staggering is noticed for \emph{N} = 50-53 for the Br isotope 
chain and for \emph{N} = 50-55 for the Rb isotope chain but is comparably smaller than for other models.

\subsection{Two neutron separation energies and the shell gap}
\label{sec:2}
Figure~9 displays the two-neutron separation energies (S$_{2n}$) of neutron-rich Br, Kr, Sr, Zr and Mo isotopes derived from the mass excess values 
measured by the JYFLTRAP and ISOLTRAP setups \cite{UH06,PA06}. The data is complemented by the values from the AME03\cite{GA03}.

\begin{figure}
\resizebox{0.5\textwidth}{!}{%
  \includegraphics{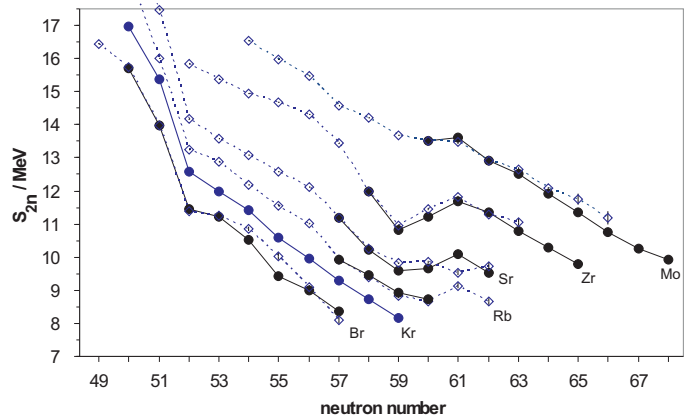}
}
\caption{Experimental two-neutron separation energy (S$_{2n}$)~plotted as a function of~neutron number. Empty diamonds connected with dashed lines 
present the values cited in the Atomic Mass Evaluation 2003 \cite{GA03}. Full circles connected with solid line indicate the new and previous 
\cite{UH06} experimental values from JYFLTRAP (Br, Rb, Sr, Zr and Mo) and ISOLTRAP (Kr) \cite{PA06}.}
\label{fig:9}       
\end{figure}

The occurrence of nuclear shell effects is clearly observed as a function of neutron number. The main shell closure at \emph{N} = 50 manifests itself 
in a steep decrease of S$_{2n}$ values for all nuclei with \emph{N} = 51 and \emph{N} = 52. A similar change of the slope is observed at the \emph{N} 
= 56 neutron subshell closure. This change of the slope is most prominent for Zr due to the fact that \emph{Z} = 40 is a semi-magic proton number. At 
\emph{N} = 59 for Zr isotopes the change in the slope of the S$_{2n}$ curve shows very clearly a sudden phase transition from a spherical to deformed 
shape of the ground state. This effect is also visible for Sr and for Rb at \emph{N} = 58 to 60. A similar trend continues in the case of the Mo 
isotopes that are less deformed than the Zr isotopes at these neutron numbers. This phenomenon has been interpreted by K. Heyde \emph{et al.} as due 
to a coexistence of different shapes at low energies \cite{KH04}. However no such conclusion can yet be drawn for the Kr and Br chains due to the lack 
of experimental data in this region.

\begin{figure}
\resizebox{0.5\textwidth}{!}{%
  \includegraphics{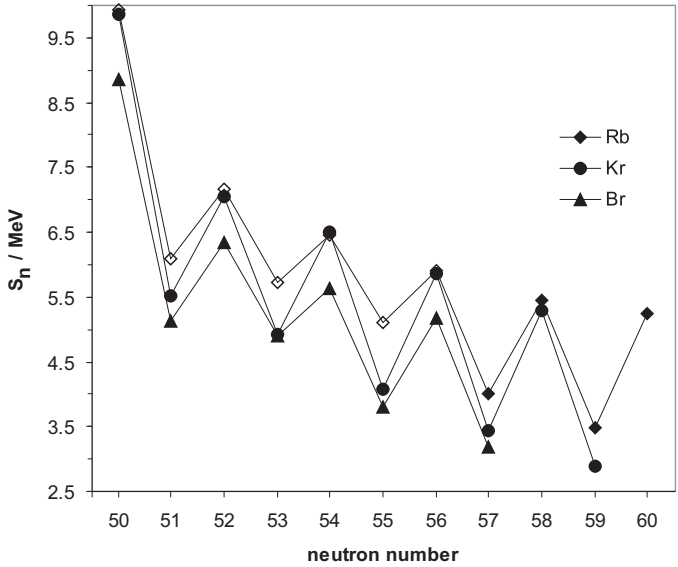}
}
\caption{One-neutron separation energy (S$_{n}$)~plotted as a function of neutron number. The filled symbols indicate the experimental values from 
JYFLTRAP (Br, Rb) and ISOLTRAP (Kr) \cite{PA06}. Empty diamonds present the values cited in the Atomic Mass Evaluation 2003 \cite{GA03}.}
\label{fig:10}       
\end{figure}

Figure~10 shows the experimental neutron separation energies for Br, Rb and Kr. The experimental neutron capture Q-value for $^{91}$Br is 3.20(1) MeV.  
The value clearly indicates that the investigated nuclei are very close to the r-process path. Our new mass values in this region will substantially 
help to improve the theoretical predictions and the network calculations of the r-process path.   

Large shell gaps are a predominant feature of magic nuclei. In the language of the shell model, such magic-number nuclei have
a closed-shell structure leading to their pronounced stability. This phenomenon is well understood for the ground states of nuclei close to the valley 
of $\beta$-stability. However, the situation seems to be quite different for highly unstable nuclei. One example is that of abnormal neutron to proton 
number ratios. Several theoretical calculations and experimental results have shown indications of a quenching of the $\emph{N}_{0}$ = 20 shell in 
neutron-rich isotopes \cite{YU99,TM95,ZD06}. Also the $\emph{N}_{0}$ = 50 shell gap has been predicted by different mass models \cite{MS02,RCN99} to 
be reduced towards $\emph{Z} \sim 28$. This phenomenon has been explained by refs. \cite{YU99,MS83} and can be understood from a single particle 
spectrum. In the case of nuclei far from the stability line, the single particle spectrum is wider, therefore for these nuclei the neutron pairing 
energy will be enhanced and, as a result, the shell-gaps are weakened.

The shell-gap energy ($\Delta$(\emph{N}$_{0}$)) can be calculated by using the one or two neutron separation energies  

\begin{equation}
\Delta(N_{0})_{2n} = S_{2n}(Z, N_{0}) - S_{2n}(Z, N_{0}+2)
\end{equation}

and 
\begin{equation}
\Delta(N_{0})_{1n} = S_{n}(Z, N_{0}) - S_{n}(Z, N_{0}+1)
\end{equation}

where $\emph{N}_{0}$ defines the magic neutron number. Figure~11 displays two regions of interest regarding the present work and other recent data. 
The JYFLTRAP values from this work, previous work and the new ISOL\-TRAP values for Kr are included in this plot \cite{UH06,AJ07,PA06}. Our $\Delta 
(\emph{N}_{0})_{2n}$ data for \emph{Z} = 35 at $\emph{N}_{0}$ = 50 confirm the data that showed a clear trend towards the shell quenching as predicted 
by several models. At $\emph{N}_{0}$ = 56 and \emph{Z} = 40 an enhanced shell gap energy is observed as expected because of both neutron and proton 
subshell closures. When comparing the $\emph{N}_{0}$ = 50 and $\emph{N}_{0}$ = 56 shell gaps (calculated according to Eq.~3) the effect for 
$\emph{N}_{0}$ = 56 is more pronounced. The gap is reduced by about a factor of two from \emph{Z} = 40 to \emph{Z} = 35. On the contrary, the change 
in the $\Delta (\emph{N}_{0})_{1n}$ shell gap energies are considerably smaller. For $\emph{N}_{0}$ = 56 no essentials reduction of the shell gap is 
observed down to \emph{Z} = 35. 

\begin{figure*}
\begin{center}
\resizebox{0.999\textwidth}{!}{%
  \includegraphics{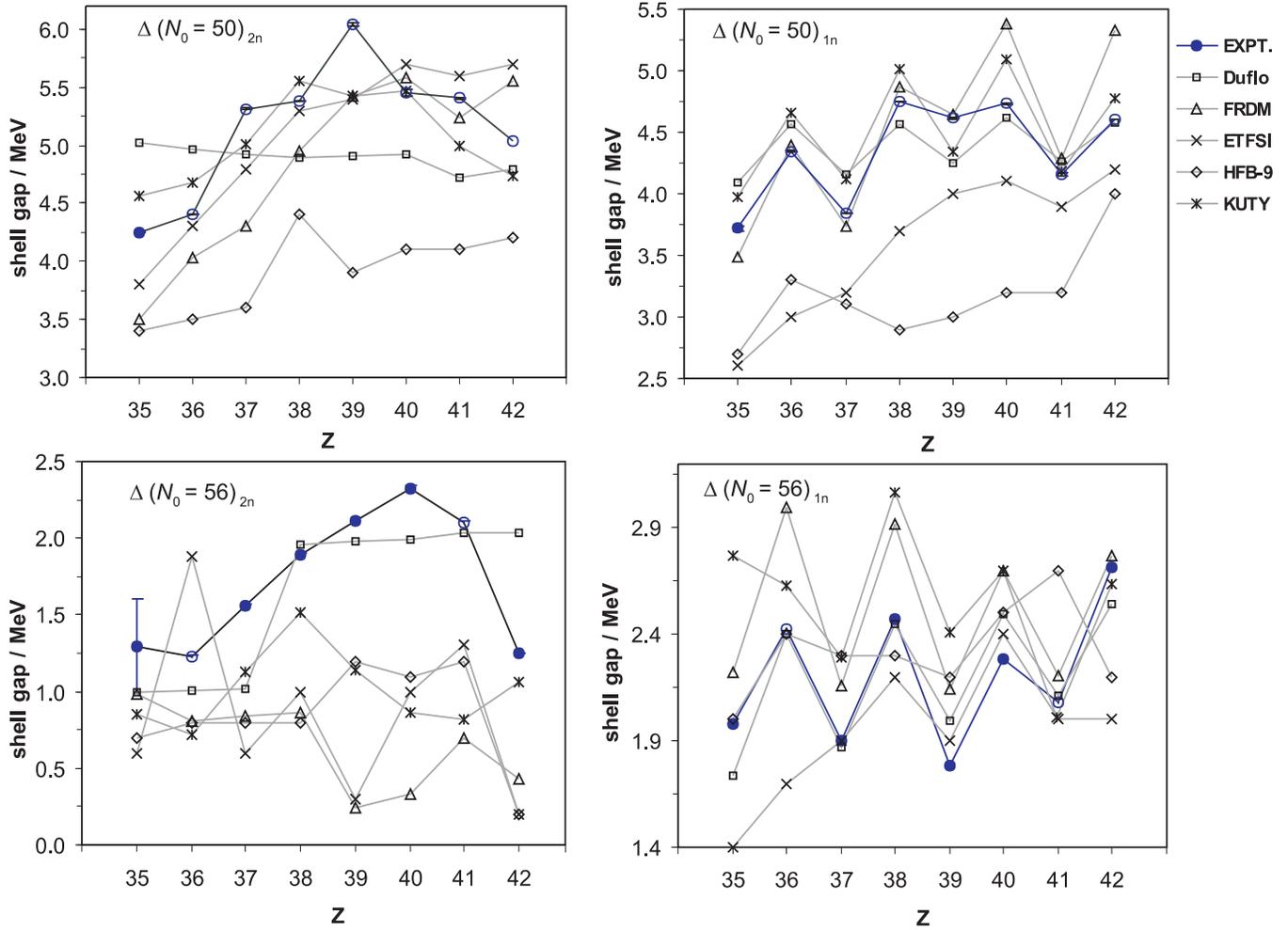}
}
\caption{Shell gap as a function of proton number for neutron numbers $\emph{N}_{0}$ = 50 and 56. Left side plots are calculated using two-neutron 
separation energies and right side plots are calculated using neutron separation energies. The filled circles indicate the JYFLTRAP values and the 
empty circles are taken from the AME03 Mass Table. The theoretical values are indicated by the symbols shown in figures.}
\label{fig:11}       
\end{center}
\end{figure*}

A large scattering of the shell gap values calculated by different models is observed for the \emph{N}$_{0}$ = 50 shell closures. The $\Delta 
(\emph{N}_{0}=50)_{2n}$ data is reasonably well reproduced by KTUY, FRDM, ETFSI models whereas no or only weak reduction is given by the Duflo and 
Zucker and HFB-9 models. None of the models can reproduced the observed shell gap energies at \emph{N} = 56. 
            
\section{Conclusions}
The masses of $^{85-92}$Br and $^{94-97}$Rb isotopes have been measured in the neighborhood of the r-process path with a relative precision of about 
10$^{-8}$. A significant deviation is observed compared to the older data for Br isotopes. The model mass excess values have an offset with the 
JYFLTRAP mass excess values of about 400 keV for the Br isotopes (see Fig.~7). On the other hand, the model mass excess values diverge with the 
increasing neutron number for the Rb isotopes (see Fig.~8). The neutron capture Q-value of $^{91}$Br indicates that this isotope might directly lie on 
the r-process path. A reduced shell gap energy is observed at \emph{Z} = 35 compared to \emph{Z} = 40. The theoretical shell gap values are in a poor 
agreement with the experimental results. 

More exotic Rb and Br isotopes will be measured in the future to gain a comprehensive understanding of the shell effects at lower \emph{Z} regions. To 
complete the shell gap plots towards the lower \emph{Z} region for $\emph{N}_{0}$ = 50, future JYFLTRAP experiment will aim to measure the masses of 
the neutron-rich isotopes further down towards \emph{Z} = 28.                
\label{sec:5} 

\begin{acknowledgement}
This work has been supported by the TRAPSPEC Joint Research Activity project under the EU 6th Framework program "Integrating Infrastructure Initiative 
- Transnational Access", Contract Number: 506065 (EURONS) and by the Academy of Finland under Finnish center of Excellence Program 2006-2011 (Nuclear 
and Accelerator Based Physics Program at JYFL and project number 202256 and 111428).
\end{acknowledgement}
 
%

\end{document}